\documentstyle[preprint,aps]{revtex}

\def\bm#1{\pmb{${#1}$}}

\begin{document}
\draft
\preprint{{\sf hep-ph/9701251} -- LPC 96/57}
\title{Neutrino-less Double Beta Decay with Composite Neutrinos.}
\author{O.~Panella$^{(a,b)}$\footnote{Corresponding author. E-mail:
o.panella@pg.infn.it}, C.~Carimalo$^{(b)}$ ,
Y.~N.~Srivastava$^{(a,c)}$ and A.~Widom$^{(c)}$}
\address{
$^{a)}$Istituto Nazionale di Fisica Nucleare, Sezione di Perugia\\
Via A.~Pascoli, I-06123 Perugia, Italy\\
$^{b)}$Laboratoire de Physique Corpusculaire, Coll\`ege de France\\
11 Place Marcelin Berthelot, F-75231, Paris Cedex 05, France\\
$^{c)}$Physics Department, Northeastern University, Boston, 
Ma. 02115 }
\date{Revised: May 1997}
\maketitle
\begin{abstract}
We study in detail the contribution of heavy composite Majorana neutrinos
to neutrino-less double beta decay ($0\nu\beta\beta$). 
Our analysis confirms the result of a previous
estimate by two of the authors. Excited
neutrinos couple to the electroweak gauge bosons through a magnetic type
effective Lagrangian.  
The relevant nuclear matrix element is related to
matrix elements available in the literature and current bounds
on the half-life of $0\nu\beta\beta$ are converted into bounds on the
compositeness scale and/or the heavy neutrino mass. Our bounds are
 of the same order
of magnitude as those available from accelerator experiments. 
\end{abstract}



\section{Introduction}

Neutrinoless double beta decay ($0\nu\beta\beta$), see Fig. 1, 
is certainly one 
of the more interesting non-accelerator processes that are presently 
being searched for.
The interest in this process stems from the fact that its observation
would undoubtedly signal lepton number violation, and at the same time
would  shed light onto the nature of the neutrino, one of the most 
elusive elementary particles. For these reasons it has received
considerable attention both from the nuclear and the particle physics 
community~\cite{trento}.

In the standard model, $0\nu\beta\beta$ can only proceed if the 
neutrino is of Majorana type and has a {\it non zero} mass.
A number of  mechanisms studied in 
models beyond the standard electroweak theory~\cite{mohapatra},
have verified that neutrino-less double beta 
decay is a very good probe of physics beyond the standard 
model. The experimental lower bound on the lifetime 
of the decay has been used to obtain constraints on the scale  of 
new physics. 

Recent work along these 
lines include: ($i$) an investigation of new super-symmetric contributions
from R-parity violating MSSM~\cite{hirsch1} shows  that
constraints on parameters of the model from  non-observation
of $0\nu\beta\beta$ are stronger than those available from
accelerator experiments;
($ii$) a detailed analysis of the contribution from left right symmetric
models\cite{hirsch2}; ($iii$) a study of the effective low energy 
charged current lepton quark interactions due to the exchange
of heavy leptoquarks~\cite{hirsch3}. The phenomenology
of Majoron models has also been studied in detail~\cite{burgess}. 

Panella and Srivastava~\cite{panella} were the first
to show that the compositeness scenario can
give an additional contribution to $0\nu\beta\beta$ and    
derived  bounds on the compositeness parameters 
from the non observation of $0\nu\beta\beta$. They explored
phenomenologically the idea that the excited state of an ordinary neutrino
might be a heavy Majorana neutral with a mass $M_N$ ranging from a few
hundred GeV up to a TeV. 
However, the nuclear aspect of the
calculation was treated only approximatively: time ordering of the
hadronic charged current was neglected and  an upper bound for the
nuclear matrix element was used in deriving  constraints on the
compositeness parameters. 
Use of the HEIDELBERG-MOSCOW $\beta\beta$ experiment 
lower bound~\cite{HM} on the half-life of the 
decay $^{76}\hbox{Ge} \to ^{76}\hbox{Se} + 2\, e^- $, yielded  
the following constraint~\footnote{the numerical value used in  
ref.~\cite{panella} was: $T^{0\nu\beta\beta}_{1/2} \ge 5.1\times 
10^{24}\hbox{yr}$.} on the scale ($\Lambda_{\hbox{c}}$), the heavy 
neutrino mass ($M_N$), and the dimensionless coupling 
constant $f$~\cite{panella}: 
\begin{equation}
\label{op}
|f| \le 3.9 
\frac{\Lambda_{\hbox{c}}}{1\, \hbox{TeV}}\biggl(\frac{M_N}{1\, 
\hbox{TeV}}\biggr)^{1/2}.
\end{equation}

Apart from the obvious desire to improve on the above mentioned
approximations the main motivation for the present work is twofold.
Firstly, after the completion of the work of ref.~\cite{panella},
appeared a related work by Takasugi~\cite{takasugi} who
considered the same problem ($0\nu\beta\beta$ via the exchange of an heavy
composite Majorana neutral) but arrived at quite different
conclusions~\cite{takasugi}~\footnote{ the numerical value used in
ref.~\cite{takasugi} was: $T^{0\nu\beta\beta}_{1/2} \ge 5.6\times
10^{24}\hbox{yr}$.}: 
\begin{equation} \label{et} |f| \le 5.8 \times 10^{-3}
\frac{\Lambda_{\hbox{c}}}{1\, \hbox{TeV}}\biggl(\frac {1\,
\hbox{TeV}}{M_N}\biggr)^{1/2}.  
\end{equation} 
In view of this discrepancy
it is, of course, mandatory to investigate further the problem in order to
understand if  reliable constraints on the compositeness scale from
$0\nu\beta\beta$ are given by Eq.~(\ref{op}) or  by Eq.~({\ref{et}).

Secondly, there is a recent claim by the CDF collaboration of a possible
signal of compositeness in high energy proton-proton
collisions~\cite{CDF}: the measured cross section for large transverse
energy jets is significantly higher than predictions based on perturbative
QCD calculations to order O($\alpha_s^3$); a compositeness scale of
$\Lambda_{\hbox{c}} = 1.6 \, \hbox{TeV}$ is suggested by the CDF
study.
Were this claim to withstand further data and analysis (such as angular
distribution of dijets presentely underway), the interest in new physics
effects arising from a composite scenario will undoubtly increase
enormously. If so, low-energy processes, such as
$0\nu\beta\beta$, could play a complementary role and hence are 
worth  further investigation. 

In this paper we present a detailed analysis of the composite Majorana
neutrino contribution to $0\nu\beta\beta$; we show that: $(i)$ the
peculiarity of the dimension five effective coupling ($\sigma_{\mu\nu}$)
shows up in giving a larger than usual importance to the high-energy
behaviour of the hadronic current correlation function; $(ii)$ the nuclear
matrix element is calculated exactly since it can be related to matrix
elements already known; $(iii)$ the results of Panella and
Srivastava~\cite{panella} remain essentially unchanged; $(iv)$ the
calculation by Takasugi presented in ref.~\cite{takasugi} is not
consistent, and its conclusions are not correct. 

\section{Effective Lagrangians for Compositeness}

The idea that at an energy scale $\Lambda_{\hbox{c}}$,  quarks and 
leptons might show an internal structure
has been around for quite some time\cite{preons}. Although many models 
describing quarks and leptons in terms of bound states of yet  more
fundamental entities (preons) have been 
proposed, so 
far, no consistent dynamical composite theory has been found.  
Phenomenologically however this idea can be probed by
observing that one natural, model independent, consequence
of compositeness  is the existence of excited states of the ordinary 
fermions with masses at least of the order of the 
compositeness scale.

We review here, for the reader's convenience, the effective
interactions used in the literature to describe possible
manifestations of lepton and quark substructure. A more
detailed discussion of compositeness phenomenology can be found
in refs.~\cite{pantrento,pdg}.

Effective couplings between the heavy and light leptons and quarks have been
proposed, using weak isospin 
($I_W$) and hyper-charge ($Y$) conservation.
Within this model, it is assumed that 
the lightness of the ordinary leptons could be related to some
global unbroken chiral symmetry which would produce massless 
bound states of preons in the absence of weak perturbations due to
$SU(2)\times U(1) $ gauge and Higgs interactions. 
The large mass of the excited leptons arises from the unknown underlying
dynamics and {\it not} from the Higgs mechanism. 

Assuming that such states are grouped in $SU(2) \times U(1)$ 
multiplets, since  light  fermions have $I_W=0,1/2$
and electroweak  gauge bosons have $I_W=0,1$,
only
multiplets with $I_W \leq 3/2$ can be excited in 
the lowest order perturbation theory.   
Also, since none of the gauge fields carry hyper-charge, a
given excited multiplet can couple only to a light multiplet with 
the same $Y$. 

In addition, conservation of the electro-magnetic
current forces the transition coupling of heavy-to-light fermions
to be of the magnetic moment type respect to any electroweak
gauge bosons~\cite{cab}.
In fact, a $\gamma_{\mu}$ transition coupling between $e$ and  $e^*$
mediated by the ${\vec W}^{\mu}$ and $B^{\mu}$ gauge fields,
would result in an electro-magnetic current of the type $j^{\mu}_{e.m.}
\approx \bar{\psi}_{e^*} \gamma^{\mu} \psi_e$ which would not be 
conserved due to the different masses of excited and ordinary
fermions, (actually it is expected that $m_{e^*} \gg m_e$).

Let us here restrict to the first family and consider 
spin-$1/2$ excited states grouped in multiplets with 
$I_W=1/2 $ and $ Y=-1$ (the so called homodoublet model~\cite{pdg}), 
\begin{equation}
L= {N \choose E} 
\end{equation} 
which can couple to the light  left-handed multiplet
\begin{equation}
\ell_L = {\nu_L \choose e_L} ={{1-\gamma_5} \over 2}
{\nu \choose e}
\end{equation}
through the gauge fields ${\vec W}^{\mu}\, \hbox{and} B^{\mu}$.
The relevant interaction can be written\cite{cab} 
in terms of two {\it new} independent coupling constants $f$ and $f'$:
\begin{eqnarray}
{\cal L}_{int}& = & \frac{gf}{\Lambda_{\hbox{c}}} \bar{L}\sigma_{\mu\nu}
\frac{\vec\tau}{2} l_L \cdot \partial^\nu\vec{W}^\mu \cr
& & \phantom{xxxxx}
+\frac{g'f'}{\Lambda_{\hbox{c}}}
\biggl(-\frac{1}{2} \bar{L}\sigma_{\mu\nu}
l_L \biggr)\cdot \partial^\nu B^\mu + \hbox{h.c.}
\end{eqnarray}
where ${\vec \tau}$ are the Pauli $SU(2)$
matrices, $g$ and $g'$ are the usual $SU(2)$ and $U(1)$ gauge coupling
constants, and the factor of $-1/2$ in the second term is the
hyper-charge of the $U(1)$ current. 
This effective Lagrangian has been widely used in the literature
to predict production cross sections and decay rates of the excited
particles~\cite{pdg,ruj,baur}.

The extension to quarks and strong interactions as well as to
other multiplets and a detailed discussion of the spectroscopy 
of the excited particles can be found in the literature~\cite{pan}.

Here, let us  write down explicitly the interaction Lagrangian 
describing the  coupling of
the heavy excited neutrino with the light electron, which  will 
be slightly generalized 
in the following section in order to discuss bounds on the 
compositeness effective couplings from low-energy, nuclear, 
double-beta decay: 
\begin{equation}
\label{leff}
{\cal L}_{eff} =  \bigl({ g f \over {\sqrt 2} \Lambda_{\hbox{c}}} \bigr)
\Bigl\{ 
\Bigl( {\overline N}
\sigma^{\mu \nu} {1-\gamma_5 \over 2} \, e \Bigr)
\, \partial_{\nu} W_{\mu}^+\Bigl\}\, + \, \hbox{h.c.} 
\end{equation}

\section{Neutrino-less Double Beta Decay (\bm{0\nu\beta\beta}).}

The transition amplitude of $0\nu \beta \beta$ decay is calculated 
according to the interaction Lagrangian:
\begin{eqnarray}
{\cal L}_{int}& =& {g \over  \sqrt{2}}\bigg\{
{f \over \Lambda_{\hbox{c}}}
\bar{\psi}_e(x) \sigma_{\mu\nu}(\eta_L L +\eta_R R)\psi_N(x)
\partial^\mu W^{\nu (-)}(x)  \nonumber \\
& \qquad & \qquad \qquad  +\cos\theta_C J^h_\mu (x) 
W^{\mu (-)}(x) + h.c. \biggr\}
\end{eqnarray}
where we have generalized the interaction in Eq.~(\ref{leff})
to include right-handed couplings (in order to allow comparison
with other models than the homo-doublet one), although we 
will 
assume chirality conservation i.e. $(\eta_L,\eta_R) = (1,0)\, \hbox{or} \, 
(0,1)$; $R=(1+\gamma_5)/2$, $L=(1+\gamma_5)/2$, $\theta_C$ is the 
Cabibbo angle 
($\cos\theta_C = 
0.974$ ) and $J_\mu^h$ is the hadronic weak charged current.

We have:          
\begin{eqnarray}
S_{fi}& =& (cos\theta_C)^2 ({g \over 2 \sqrt{2}})^4 
\left( {f \over \Lambda_{\hbox{c}}}\right)^2 ({1\over 2})
\int\, {d^4q \over (2\pi)^4}\, d^4x\, d^4y\,  \exp [-iq\cdot (x-y)]
\times \nonumber \\
&\, & 4\,\times {1\over \sqrt{2}}(1-P_{12})
{\bar \psi}(p_2)\sigma_{\mu\lambda}(\eta_L L+\eta_R R)
{\not\! q +M_N \over q^2 -M_N^2} (\eta_L L +\eta_R R)
\sigma_{\nu\rho}\psi^C(p_1) \times
\nonumber \\
&\ &  \exp[i(p_1\cdot x +p_2 \cdot y)]\,(q-p_1)^\lambda(q+p_2)^\rho 
{\langle F|T\bigl[ J^\mu_h(x) \ J^\nu_h(y)\bigr]|I\rangle
\over [(q-p_1)^2 -M_W^2] [(q+p_2)^2-M_W^2]}
\label{amplitude}
\end{eqnarray}
where   
$(1-P_{12})/\sqrt{2}$ is the antisymmetric operator due to 
the production of two identical fermions.
Now let us change variables of integration:
\begin{equation}
\left.
\begin{tabular}{l}
$x= z+\frac{u}{2} $\\
$y= z-\frac{u}{2} $
\end{tabular}
\right\} \quad \hbox{with}\quad 
d^4x\, d^4y=d^4z\, d^4u .
\end{equation}
In addition we make the {\it ansatz} that the hadronic  current
be given by the corresponding sum of the nucleonic
charged current:
\begin{equation}
J_\mu^h(x) =\sum_i J^{(i)}_\mu(x) .
\end{equation}
This implies:
\begin{equation}
\langle F|T\bigl[ J^\mu_h(x) \ J^\nu_h(y)\bigr]|I\rangle
= \exp [i(p_F-p_I)\cdot y)] \,
\langle F|T\bigl[ J^\mu_h(x-y) \ J^\nu_h(0)\bigr]|I\rangle
\end{equation}
We have:
\begin{itemize}
\item $p_1\cdot x +p_2\cdot y \, = \, (p_1+p_2)\cdot z +(p_1-p_2)\cdot u/2 $
\item $(\eta_L L+\eta_R R){(\not\! q +M_N )} (\eta_L L +\eta_R R) = 
(\eta_L^2 L +\eta_R^2 R) M_N$\ \ \ \ \  if \ \  $\eta_R\eta_L =0$.
\item  
$(g/(2\sqrt{2}))^4 =M_W^4G_F^2/2 $
\end{itemize}
Thus  we arrive at:
\begin{eqnarray}
S_{fi}& =& (cos\theta_C)^2 G_F^2 
\left( {f \over \Lambda_{\hbox{c}}}\right)^2 
\int\, {d^4q \over (2\pi)^4}\, d^4z\, d^4u\, \exp(-iq\cdot u)
\times \nonumber \\
&\, &M_N\, M_W^4 \, {1\over \sqrt{2}}(1-P_{12})
{\bar \psi}(p_2)\sigma_{\mu\lambda}
\sigma_{\nu\rho}(\eta_L^2 L +\eta_R^2 R)\psi^C(p_1) \times
\nonumber \\
&\ &  \exp[i\, z \cdot (p_1 +p_2 +p_F -p_I)]\,
\exp \left[i\, (u/2)
\cdot (p_1 -p_2 -p_F +p_I)\right]\,\times \nonumber \\
&\ &(q-p_1)^\lambda(q+p_2)^\rho 
{\langle F|T\bigl[ J^\mu_h(u) \ J^\nu_h(0)\bigr]|I\rangle
\over [(q-p_1)^2 -M_W^2] [(q+p_2)^2-M_W^2]( q^2 -M_N^2)}
\end{eqnarray}
The integration over $d^4z$ gives the energy-momentum
conservation and if we define:
\begin{itemize}
\item $S_{fi} =(2\pi)^4\delta^4(p_I-p_F-p_1-p_2) \, T_{fi} $
\item $G_{eff} =\cos\theta_C \, G_F\, (f/\Lambda_{\hbox{c}}) $
\end{itemize}
we obtain:
\begin{eqnarray}
T_{fi}& =& \frac{G_{eff}^2}{\sqrt{2}}\, (1-P_{12})
\int\, {d^4q \over (2\pi)^4}\, d^4u \, \exp[-i(q-p_1)\cdot u]
\times \nonumber \\
&\, &M_N\, M_W^4 \,
{\bar \psi}(p_2)\sigma_{\mu\lambda}
\sigma_{\nu\rho}(\eta_L^2 L +\eta_R^2 R)\psi^C(p_1)
\,(q-p_1)^\lambda(q+p_2)^\rho  \times
\nonumber \\
&\ & 
{\langle F|T\bigl[ J^\mu_h(u) \ J^\nu_h(0)\bigr]|I\rangle
\over [(q-p_1)^2 -M_W^2] [(q+p_2)^2-M_W^2]( q^2 -M_N^2)}
\end{eqnarray}
Next, we neglect $p_1$ and $p_2$ everywhere with respect to $q$ except
in the electronic  wave functions. Using  the identity:
\begin{eqnarray}
\label{spin_iden}
& &(1-P_{12})\,{\bar \psi}(p_2)\sigma_{\mu\lambda}
\sigma_{\nu\rho}(\eta_L^2 L +\eta_R^2 R)\psi^C(p_1)
={\bar \psi}(p_2)\bigl\{\sigma_{\mu\lambda},
\sigma_{\nu\rho}\bigr\}(\eta_L^2 L +\eta_R^2 R)\psi^C(p_1)\cr
&\phantom{xx}& = 2\, {\bar \psi}(p_2)
(\eta_{\mu\nu}\eta_{\lambda\rho} -\eta_{\mu\rho}\eta_{\lambda\nu}
+i\gamma_5\epsilon_{\mu\lambda\nu\rho})
(\eta_L^2 L +\eta_R^2 R)\psi^C(p_1)
\end{eqnarray}
we find:
\begin{equation}
\label{tfi_zero}
T_{fi} = \frac{2G_{eff}^2}{\sqrt{2}}\,
{\bar \psi}(p_2)(\eta_L^2 L +\eta_R^2 R)\psi^C(p_1)
\, M_N M_W^4 \, \int \frac{d^4q}{(2\pi)^4}\,
\frac{(q^2\eta^{\mu\nu} -q^\mu q^\nu) \, W_{\mu\nu}(q)}
{(q^2-M_W^2 +i\epsilon)^2(q^2-M_N^2+i\epsilon)} \, \, ,
\end{equation}
where we have defined:
\begin{equation}
W^{\mu\nu}(q) = \int d^4x \, \exp(-iq\cdot x) \,
\langle F|T\bigl[ J^\mu_h(x) \ J^\nu_h(0)\bigr]|I\rangle.
\end{equation}
Eq.~(\ref{tfi_zero}) gives:
\begin{eqnarray}
\label{tfi_one}
T_{fi} &=& \frac{2G_{eff}^2}{\sqrt{2}}\,
{\bar \psi}(p_2)(\eta_L^2 L +\eta_R^2 R)\psi^C(p_1)
\, M_N M_W^4 \, \int \frac{d^3\bm{q}}{(2\pi)^3}\,\cr
& \times & \int \frac{dq_0}{2\pi}
\frac{-q^iq^j\bigl[W_{ij}-\eta_{ij}(W_{00}+W_k^k)
\bigr] +q_0^2\, W_k^k}{(q_0^2 -\omega_N^2+i\epsilon)\,
(q_0^2 -\omega_W^2+i\epsilon)^2}.
\end{eqnarray}
where we have defined: $\omega_{W,N} = \sqrt{\bm{q}^2 +M_{W,N}^2}$
and terms in $q_0q_i$ have been dropped because they do not contribute
to $0^+ \to 0^+ $ transitions.

Inserting a complete set of intermediate states one can cast $W_{\mu\nu}$
in the form:
\begin{equation}
W_{\mu\nu}(q)=
(-i)\, \int d^3\bm{x} \, \exp(i\bm{q}\cdot \bm{x})\, 
\sum_X\biggl\{
\frac{\langle F|J_\mu^h(\bm{x})|X\rangle\langle X| J_\nu^h(0)|I\rangle}
{q_0-E_F+E_X-i\epsilon}
+
\frac{\langle F|J_\nu^h(\bm{0})|X\rangle\langle X| J_\mu^h(\bm{x})|I\rangle}
{-q_0-E_I+E_X-i\epsilon}\biggr\}
\end{equation}

Using some known results  on matrix elements of one particle
operators, the quantity $W_{\mu\nu}(q)$ can be readily evaluated.
 Note that the sum over intermediate states
includes an integration over the continuous part of the spectrum,
namely the center of mass momentum $\bm{P}$:
$\sum_X \to (2\pi)^{-3}\int d^3\bm{P}\, \sum_n$. This integration
can be carried out analytically.

The energy of a state $|X\rangle $ is $E_X =E_{CM}(\bm{P}) +\epsilon_n$,
where  $E_{CM}(\bm{P})$ is the energy of the center of mass 
translational motion and $\epsilon_n$ is the 
excitation energy~\cite{lipkin}.
($E_{CM}(\bm{P}) = \sqrt{\bm{P}^2 + (m_pA)^2} $, with
$m_p$ the proton mass and $A$ the mass number of the nucleus).

In the closure approximation (routinely applied in double beta 
calculations), the excitation energy $\epsilon_n $
of the intermediate state is replaced by an average value
$ \bar{\epsilon}_n $ and the sum over the discrete part
of the intermediate states is performed. 

The center of mass motion 
($\bm{R} =({1}/{A}) \sum_i\bm{r}_i $ ) can be separated out
so that we have~\cite{lipkin}:
\begin{equation}
\bm{\xi}_i=\bm{r}_i-\bm{R}\qquad
\sum_i\bm{\xi}_i = \bm{0}
\end{equation}
In terms of the new coordinates ($ \bm{\xi}_i $) we have:
\begin{equation}
\psi(\bm{r}_1,......,\bm{r}_A)
= \exp\left( i\,\bm{P}\cdot\bm{R}\right)\,
\Phi(\bm{\xi}_1,......,\bm{\xi}_{A-1})
\end{equation}
Using this for the calculation of one body 
operator matrix elements~\cite{lipkin} one obtains
(for states of definite center of mass momentum):
\begin{equation}
\langle F | O(\bm{r}) |X\rangle= \sum_i
\langle\langle F|\, \exp[i\,(\bm{P}_X -\bm{P}_F) \cdot (\bm{r} -\bm{\xi}_i)] 
\, O^{(i)} (\bm{P}_X -\bm{P}_F)\,|X\rangle\rangle
\end{equation}
where we denote with $\langle \langle ...\rangle \rangle $ the 
matrix element in the space of the $A-1$ relative coordinates.

After integrating over  the center of mass motion, 
we obtain:
\begin{equation}
W_{\mu\nu} (q) = \frac{2i\Delta}{q_0^2-\Delta^2 +i\epsilon}
\langle\langle F | \sum_{kl}\,  \exp(i\bm{q}\cdot \bm{\xi}_{kl}) \,
j_\mu^{(k)}(-\bm{q})\, j_\mu^{(l)}(\bm{q})| I \rangle \rangle 
\end{equation}
(in the limit 
$\bm{p}_F \approx \bm{p}_I \approx 0 $).  
We have also used the approximation $\epsilon_F \approx \epsilon_I$ :
\begin{equation}
\Delta = \bar{\epsilon}_n -\epsilon_I +E_{CM}(\bm{q}) -E_{CM}(\bm{0})
\end{equation}
Since  on the average $|\bm{q}| \approx
40 $ MeV, we may conclude that the center of mass motion does not
give any appreciable contribution to $\Delta$:
\begin{equation}
\Delta \approx \bar{\epsilon}_n -\epsilon_I \approx 10 \, \, \hbox{MeV}
\end{equation}
Next, we use the non-relativistic
limit for the nuclear current~\cite{deshalit} : 
\begin{equation}
j_\mu^{(k)} (\bm{q}) = f_A(\bm{q}^2)\times \, \left\{
\begin{tabular}{ll}
$\phantom{-}g_V\tau_+^{(k)}$ & if $\mu=0$\\
$-g_A{\tau_+}^{(k)} (\sigma_k)_i$ & if $\mu=i$
\end{tabular}
\right.
\end{equation}
where $\vec{\sigma}_k $ is the spin matrix of the $\bm{k}$-th nucleon, 
and a nucleon form factor,
\begin{equation}
f_A(\bm{q}^2)={1\over (1+\bm{q}^2/ m_A^2)^2}
\end{equation}
with $m_A=0.85 $ GeV, 
has been introduced to account for the finite size of the nucleon.
The latter  is known to give a sizable contribution 
for the heavy neutrino case. 

Within the above approximations, we are thus led to:
\begin{eqnarray}
\label{naive}
W_{0,0} (q) &=& \frac{2i\Delta}{q_0^2-\Delta^2 +i\epsilon} \,\, g_V^2 \,
f_A^2(\bm{q}^2) \, 
\langle\langle F | \sum_{kl}\,  \exp(i\bm{q}\cdot \bm{r}_{kl}) \,
\tau_+^{(k)}\tau_+^{(l)} | I \rangle \rangle \cr
W_{i,j} (q) &=& \frac{2i\Delta}{q_0^2-\Delta^2 +i\epsilon}\,\, g_A^2 \,
f_A^2(\bm{q}^2) \,
\langle\langle F | \sum_{kl}\,  \exp(i\bm{q}\cdot \bm{r}_{kl}) \,
\tau_+^{(k)}\tau_+^{(l)}(\sigma_k)_i(\sigma_l)_j | I \rangle \rangle 
\end{eqnarray}

These expressions for $W_{\mu\nu}$ are standard and the closure
approximation has been routinely applied in double beta 
calculations~\cite{burgess}.

\section{Decay Formulae and Nuclear Matrix Element}

We can now proceed to  calculate  the
$0\nu\beta\beta$ decay amplitude using the results of the 
previous section in Eq.~(\ref{tfi_one}). 

Defining:
\begin{eqnarray}
I(\bm{q^2}) & = & -2  \,\frac{\partial}{\partial \omega_W^2} 
\, J(\bm{q^2}) \cr
J(\bm{q^2}) & = &  \int \frac{dq_0}{2\pi i}\, 
\biggl[\frac{\Delta}{q_0^2-\Delta^2 +i\epsilon} 
%
%
\biggr]\, \frac{1}
{(q_0^2-\omega_N^2 +i\epsilon)
\,(q_0^2-\omega_W^2 +i\epsilon)} \cr
I'(\bm{q^2}) & = & -2  \,
\frac{\partial}{\partial \omega_W^2} \, J'(\bm{q^2}) \cr
J'(\bm{q^2}) & = &  \int \frac{dq_0}{2\pi i} \,
\biggl[ \frac{\Delta}{q_0^2-\Delta^2 +i\epsilon}
%
%
\bigg]\, \frac{q_0^2}
{(q_0^2-\omega_N^2 +i\epsilon)
\,(q_0^2-\omega_W^2 +i\epsilon)}
\label{q_0integ}
\end{eqnarray}
we can write:
\begin{eqnarray}
\label{tfi_last}
T_{fi} &=& \frac{2G_{eff}^2}{\sqrt{2}}\,
{\bar \psi}(p_2)(\eta_L^2 L +\eta_R^2 R)\psi^C(p_1)
\, M_N M_W^4 \, \int \frac{d^3\bm{q}}{(2\pi)^3}\, \times \cr
& & \langle \langle F | \, 
\sum_{kl} \exp{(i\bm{q}\cdot\bm{r}_{kl})}\, \tau_+^{(k)}\tau_+^{(l)}
\biggl\{ -q^iq^j \biggl[ g_A^2 \sigma_i^{(k)}\sigma_j^{(l)}
-\eta_{ij} (g_V^2 -g_A^2 \bm{\sigma}_{(k)}\cdot\bm{\sigma}_{(l)})\biggr]\, 
I(\bm{q^2}) \cr
& & - g_A^2 \bm{\sigma}_{(k)}\cdot\bm{\sigma}_{(l)} \, I'(\bm{q^2})\biggr\}
| I \rangle\rangle \times f_A^2(\bm{q}^2)
\end{eqnarray}
The integrals $J$ and $J'$ are:
\begin{eqnarray}
\label{integrals}
J &=&-\frac{1}{2}\, \frac{\Delta + \omega_W +\omega_N}{
\omega_W \omega_N \,(\omega_W +\omega_N)
(\Delta +\omega_W) (\Delta +\omega_N)}
%
%
\cr 
J' &=&\frac{1}{2} \, \frac{\Delta}  
{(\omega_W +\omega_N)(\Delta +\omega_W) (\Delta +\omega_N)}
%
%
\end{eqnarray}

We note that 
between the
three scales $\Delta$, $\omega_W  $ and $\omega_N$ involved in the
problem we have the following ordering:
\begin{equation}
\Delta \ll \omega_W \ll \omega_N
\end{equation} 
Using Eq.~(\ref{integrals}) one finds:
\begin{eqnarray}
I &\approx& - \frac{1}{\omega_W^4\omega_N^2}
\biggl[1+\frac{3}{2}\frac{\Delta}{\omega_W}+
O\biggl({\Delta^2}/{\omega_W^2}\biggr)\biggr]\cr
I' &\approx& - \frac{\Delta \omega_W}{\omega_W^4\omega_N^2}
\biggl[ -\frac{1}{2}+\frac{\Delta}{\omega_W}
+ O\biggl({\Delta^2}/{\omega_W^2}\biggr)\biggr]
\end{eqnarray} 
Using these results into Eq.~(\ref{tfi_last}) we find:
\begin{eqnarray}
\label{t_fi_D}
T_{fi} &=& \frac{2G_{eff}^2}{\sqrt{2}M_N}\,
{\bar \psi}(p_2)(\eta_L^2 L +\eta_R^2 R)\psi^C(p_1)\, g_A^2
\,  \times \cr & \phantom{} & \langle \langle F | \, 
\sum_{kl} \, \tau_+^{(k)}\tau_+^{(l)}
\biggl\{ D_{ij}(r_{kl})\biggl[ \sigma_i^{(k)}\sigma_j^{(l)}
-\eta_{ij} \biggl(\frac{g_V^2}{g_A^2} -
\bm{\sigma}_{(k)}\cdot\bm{\sigma}_{(l)}\biggr)\biggr]
-\frac{1}{2}D(r_{kl}) \bm{\sigma}^{(k)}\cdot\bm{\sigma}^{(l)}\biggr\}
| I \rangle\rangle \nonumber \\ 
\end{eqnarray}
where $ D_{ij}(r_{kl})$ and $ D(r_{kl})$ are given as:
\begin{eqnarray}
D_{ij}(r_{kl}) &=& \int \frac{d^3\bm{q}}{(2\pi)^3}\, 
\, \exp{(i\bm{q}\cdot\bm{r}_{kl})}\, \frac{q_iq_j}{(1+\bm{q}^2/m_A^2)^4} \cr
&& \cr
D(r_{kl}) &=& \int \frac{d^3\bm{q}}{(2\pi)^3}\, 
\, \exp{(i\bm{q}\cdot\bm{r}_{kl})}\, \frac{\Delta M_W}
{(1+\bm{q}^2/m_A^2)^4}\, \, . 
\end{eqnarray}
Explicitly :
\begin{eqnarray}
\label{dfunc}
D_{ij}(r_{kl}) &=& \frac{m_A^4}{4\pi R_0}\, \frac{R_0}{r_{kl}}\,
\biggl[ \delta_{ij}\, F_A(x_A) - \frac{(\bm{r}_{kl})_i\, (\bm{r}_{kl})_j}
{r_{kl}^2}
\, F_B(x_A)\biggr] \cr &&\cr
D(r_{kl}) & = & \frac{\Delta M_W m_A^2}{4\pi R_0} \,\frac{R_0}{r_{kl}}\, F_N(x_A)
\end{eqnarray}
where $R_0 = r_0A^{1/3} $ ($r_0=1.1 fm$) , $x_A =m_A r_{kl}$ and :
\begin{eqnarray}
F_A(x) &=& \frac{1}{48} \,\exp(-x)\, (x^2+x)\cr
& &\cr 
F_B(x) &=& \frac{1}{48} \,\exp(-x)\, x^3\cr
& &\cr 
F_N(x) &=& \frac{1}{48} \,\exp(-x)\, (x^3 +3x^2 +3x)
\end{eqnarray}
Inserting the results of Eq.(\ref{dfunc}) into Eq.~(\ref{t_fi_D}), one can 
cast $T_{fi}$ in the form: 
\begin{equation}
\label{tfi_last_last}
T_{fi} = \frac{2G_{eff}^2}{\sqrt{2} M_N}\,
{\bar \psi}(p_2)(\eta_L^2 L +\eta_R^2 R)\psi^C(p_1)\,
\frac{g_A^2 m_A^4}{4\pi R_0} \,  {\cal M}_{FI}
\end{equation}
where ${\cal M}_{FI}$ is :
\begin{eqnarray}
\label{mfi}
{\cal M}_{FI} & = & \langle \langle F | \, 
\sum_{kl} \, \tau_+^{(k)}\tau_+^{(l)}\, \left(\frac{R_0}{r_{kl}}\right)\,
\biggl\{ \biggl(\frac{g_V^2}{g_A^2} - \frac{2}{3}\,
\bm{\sigma}^{(k)}\cdot\bm{\sigma}^{(l)}\biggr) (3F_A -F_B) \cr
&&\phantom{xxx} -\frac{1}{3} \biggl(3\, \frac{\bm{\sigma}_{(k)}
\cdot \bm{r}_{kl} \, \bm{\sigma}_{(l)} \cdot \bm{r}_{kl}}{r_{kl}^2}
-\bm{\sigma}^{(k)}\cdot\bm{\sigma}^{(l)}\biggr) F_B
-\frac{1}{2}\frac{\Delta M_W}{m_A^2}
\, \bm{\sigma}^{(k)}\cdot\bm{\sigma}^{(l)}\, F_N
\biggr\} | I \rangle \rangle .
\end{eqnarray}
In Eq.~(\ref{mfi}) we have written  ${\cal M}_{FI}$  in terms of 
the following 
{\it known} nuclear structure matrix elements\cite{hirsch}\footnote{
Note that in ref.~\cite{hirsch} a slightly different notation is used:
$F_4= 3F_A -F_B $ and $F_5 = F_B$.}  :
\begin{eqnarray}
{\cal M}_{GT,N} &=& 
\langle \langle  F | \sum_{kl} \tau_{+}^{(k)} \tau_{+}^{(l)}
                     \bm{\sigma}_{(k)} \cdot \bm{\sigma}_{(l)}
                     \left(\frac{R_0}{r_{kl}}\right)
                      F_{N}(x_{A})
| I \rangle\rangle\cr
{\cal M}_{F,N} &=& 
\langle \langle  F | \sum_{kl} \tau_{+}^{(k)} \tau_{+}^{(l)}
                     \left(\frac{R_0}{r_{kl}}\right)
                      F_{N}(x_{A})
| I \rangle\rangle\cr
{\cal M}_{GT'} &=& 
\langle \langle  F | \sum_{kl} \tau_{+}^{(k)} \tau_{+}^{(l)}
                     \bm{\sigma}_{(k)} \cdot \bm{\sigma}_{(l)}
                     \left(\frac{R_0}{r_{kl}}\right)
                     \left[3F_A(x_A) -F_B(x_A)\right]
| I \rangle\rangle \cr
{\cal M}_{F'} &=& 
\langle \langle  F | \sum_{kl} \tau_{+}^{(k)} \tau_{+}^{(l)}
                     \left(\frac{R_0}{r_{kl}}\right)
                     \left[3F_A(x_A) -F_B(x_A)\right]
| I \rangle\rangle \cr
{\cal M}_{T'} &=& 
\langle \langle  F | \sum_{kl} \tau_{+}^{(k)} \tau_{+}^{(l)}
                     \left(\frac{R_0}{r_{kl}}\right)
                     \biggl(3\, \frac{\bm{\sigma}_{(k)}
                     \cdot \bm{r}_{kl} \, \bm{\sigma}_{(l)} 
                     \cdot \bm{r}_{kl}}{r_{kl}^2}-\bm{\sigma}^{(k)}
                     \cdot\bm{\sigma}^{(l)}\biggr)
                     \, F_B(x_A)
| I \rangle\rangle
\label{matrix_list}
\end{eqnarray}
The numerical values  for the case of $^{76}\hbox{Ge}$ 
quoted in ref.~\cite{hirsch} are given in Table I for
easy reference. 
We can decompose ${\cal M}_{FI}$ as :
\begin{eqnarray}
{\cal M}_{FI}& = &{\cal M}_{FI}^{(0)} +{\cal M}_{FI}^{(\Delta)}\cr
{\cal M}_{FI}^{(0)}& = & \frac{g_V^2}{g_A^2}\, {\cal M}_{F'}
               -\frac{2}{3}\,{\cal M}_{GT'}
               -\frac{1}{3}\,{\cal M}_{T'}\cr 
{\cal M}_{FI}^{(\Delta)}& = &-\frac{1}{2}
\frac{\Delta M_W}{m_A^2}\,{\cal M}_{GT,N}.
\label{mfi_last}
\end{eqnarray} 
and finally obtain :
\begin{eqnarray}
{\cal M}_{FI}^{(0)}& = & +8.15\times 10^{-3}\cr
{\cal M}_{FI}^{(\Delta)}& = & -6.27 \times 10^{-2}\cr
{\cal M}_{FI} & = & -5.45 \times 10^{-2}.
\label{mfi_num}
\end{eqnarray}
Let us note that in Eq.~(\ref{mfi}), 
the part of the nuclear operator which is independent of
$\Delta$ coincides 
exactly~\footnote{apart from a normalisation factor.} with the result of 
ref.~\cite{panella}. (Neglecting the time ordering
in the hadronic current is equivalent to the limit $\Delta \to 0$).
Our result in Eq.~(\ref{tfi_last_last}) shows the expected scaling 
$\sim 1/M_N$
with the composite neutrino mass $M_N$; exchanging a very heavy
Majorana neutrino one expects a factor $M_N \times
M_N^{-2}$ from the neutrino 
propagator and the exchange of heavier particles reduces the probability
of the decay. This behaviour does {\it not} appear in the formulae 
of ref.~\cite{takasugi}.

\section{Discussion}
Calculation of the half-life of the decay from Eq.~(\ref{tfi_last_last})
is now straightforward. We could use the results of ref.~\cite{panella}
but prefer instead to use the already well known phase space factors
given in \cite{doi} which take into account the distortion of the
electron wave functions going beyond the Rosen-Primakoff approximation.
The inverse half-life ($T_{1/2} = \log{2}\, T  $) is given by:
\begin{equation}
T^{-1}_{1/2} = \frac{1}{\log 2} \, \int \, \overline{^{\, }|T_{fi}|^2}\,
2\pi \delta(E_I -E_F -E_1-E_2)\, \frac{d^3\bm{p}_1}{(2\pi)^3 2E_1}
\frac{d^3\bm{p}_2}{(2\pi)^3 2E_2}.
\end{equation}
From Eq.~(\ref{tfi_last_last}) one has:
\begin{equation}
\overline{^{\, }|T_{fi}|^2} =
\frac{2G_{eff}^4}{M_N^2}\frac{g_A^4m_A^8}
{16\pi^2R_0^2} |{\cal M}_{FI}|^2 \sum_{\text{electron spins}}
|{\bar \psi}(p_2)(\eta_L^2 L +\eta_R^2 R)\psi^C(p_1)|^2
\end{equation}
The electronic  wave functions are given by:
\begin{equation}
\psi(p_{1,2}) = F_0(Z+2,E_{1,2}) \, u(p_{1,2})
\end{equation}
where  $F_0(Z+2,E_{1,2})$ is the well known Fermi
function describing the distortion of the electron wave 
in the Coulomb field of the nucleus.
We have then with straightforward algebra:
\begin{eqnarray}
\sum_{\text{electron spins}}& &
|{\bar \psi}(p_2)(\eta_L^2 L+\eta_R^2 R)\psi^C(p_1)|^2 = \cr
&=& F_0(Z+2,E_{1})\,F_0(Z+2,E_{2})
\sum_{\text{electron spins}}
|{\bar u}(p_2)(\eta_L^2 L +\eta_R^2 R)u^C(p_1)|^2\cr
&=&F_0(Z+2,E_{1})\,F_0(Z+2,E_{2})
\, (\eta_L^4+\eta_R^4)\, 2 \, p_1 \cdot p_2
\end{eqnarray}
As regards the phase-space integration, we adopt the standard 
notation and  define:
\begin{eqnarray}
\label{g01}
{\cal A}_{0\nu}&=& \frac{(G_F \cos\theta_C g_A)^4 m_e^9}{64\pi^5} \cr
G_{01}& = &\frac{{\cal A}_{0\nu}}{\log 2\, (m_eR_0)^2}
\int\,  \frac{2p_1E_1p_2E_2}{m_e^5} \, 
\delta(E_I -E_F -E_1-E_2)\times \cr
& &\phantom{xxxxxxxxxxxxxxxxxxxxx}
\,F_0(Z+2,E_{1})\,F_0(Z+2,E_{2})\, dE_1\,dE_2
\end{eqnarray}
The phase-space integral in Eq.~(\ref{g01}) is known in the literature
and its value, which completely takes account  of the Fermi functions, 
is $G_{01} = 6.4 \times 10^{-15} yr^{-1}$ (see ref.\cite{doi}).
As for the half-life, we can write it as:
\begin{equation}
\label{t1/2}
T_{1/2}^{-1} = \left(\frac{f}{\Lambda_{\hbox{c}}}\right)^4 \frac{m_A^8}{M_N^2}
\, |{\cal M}_{FI} |^2 \, \frac{G_{01}}{m_e^2}\,(\eta_L^4+\eta_R^4) .
\end{equation}

Experimentally, $0\nu\beta\beta$ decay has never been
observed so far and therefore we have available only 
lower bounds on the half-life $T_{1/2}^{\,\, \text{lower bound}}$:
\begin{equation}
\label{t1/2bound}
T_{1/2} > T_{1/2}^{\,\, \text{lower bound}}.
\end{equation}
We can use the constraint in Eq.~({\ref{t1/2bound}) into
Eq.~(\ref{t1/2}) in order to obtain constraints on the 
compositeness parameters:
\begin{equation}
\label{constraint}
\left|\frac{f}{\Lambda_{\hbox{c}}}\right| < M_N^{1/2}
\left(\frac{m_e^2}{m_A^8}\right)^{1/4}
\frac{\left[G_{01}\, T_{1/2}^{\, \, \text{lower bound}}
\, (\eta_L^4+\eta_R^4)\right]^{-1/4}}
{ |{\cal M}_{FI} |^{1/2} }
\end{equation}
Let us now consider the neutrino-less double beta decay: 
\begin{equation}
^{76}\hbox{Ge} \to ^{76}\hbox{Se} + 2\, e^- 
\end{equation}
for which the HEIDELBERG-MOSCOW $\beta\beta$ experiment provides
the current lower bound on the half-life~\cite{HMnew}:
\begin{equation}
T^{0\nu\beta\beta}_{1/2} >  7.4\times 
10^{24}\hbox{yr} 
\end{equation}
Substituting this value and the other numerical constants 
($\eta_L=1$, $\eta_R=0$)
into Eq.~(\ref{constraint}) we obtain finally:
\begin{equation}
\label{constraint_num}
|f| \le 9.06
\frac{\Lambda_{\hbox{c}}}{1\, \hbox{TeV}}\biggl(\frac{M_N}{1\, 
\hbox{TeV}}\biggr)^{1/2}.
\end{equation}

Eqs.~\ref{constraint} \& \ref{constraint_num} 
are the central results of this work.
They describe the constraints on compositeness parameters imposed by
the non observation of neutrino-less double beta decay.
A few comments are in order.

We have derived, according to the closure approximation, an expression
for the nuclear matrix element of the form ${\cal M}_{FI} = 
{\cal M}_{FI}^{(0)}+{\cal M}_{FI}^{(\Delta)}$ where ${\cal M}_{FI}^{0}$
coincides with the matrix element given in 
ref.~\cite{panella,pantrento}, and ${\cal M}_{FI}^{\Delta}$ is the 
first order correction in powers of $\Delta $. Thus in the limit
 $\Delta \to 0 $ we recover the previously published 
result~\cite{panella,pantrento}, obtained neglecting the time
ordering of the hadronic charged current c.f. Eq.~\ref{amplitude}.

Another important improvement of the present work with respect 
to~\cite{panella,pantrento} is  that the matrix element has
now been expressed in terms of {\it known} matrix 
elements~\cite{hirsch}, so that
while previously only a bound  $|{\cal M}_{FI}^{(0)}| < 0.34 $ could be
given,  we have now a definite number for ${\cal M}_{FI}$ 
c.f.~Eq.~\ref{mfi_num}. Thus while in~\cite{panella,pantrento} only
a {\it most stringent bound } for the compositeness parameters 
could be quoted, here we have derived {\it the bounds} imposed by 
$0\nu\beta\beta$ non observation. 
For definiteness we have shown in Fig.~\ref{fig:2}
 the present bounds on
$|f|$ for ($\Lambda_{\hbox{c}} = 1\, \hbox{TeV}$) as function
of the heavy neutrino mass $M_N$. The bound (solid line) is compared
with the most stringent bounds quoted in~\cite{panella,pantrento}.
We can see that the true bound is, as expected, somewhat weaker
than the previously published one.
The same conclusion can be drawn from Fig.~\ref{fig:3}
where we give, as a function of  $M_N$, the lower bound on 
$\Lambda_{\hbox{c}}$ for ($|f| =1$).

The fact that within the closure
approximation we obtain an analytic expression of the nuclear 
operator identical to the previous one (c.f.~\cite{panella,pantrento})
up to a term linear in $\Delta$, which turns out to be numerically
non negligible, gives us confidence on the correctness
of our result, namely Eq.~(\ref{mfi_last}) and Eq.~(\ref{constraint}). 

We believe that the analysis given by Takasugi~\cite{takasugi},
does not 
support his conclusion given in  Eq.~\ref{et}. That result  is 
in contradiction with the physical expectation that 
exchanging an heavier Majorana neutral reduces the probability
of the neutrino-less double beta decay and thus must impose
weaker bounds on the remaining parameters. 
We give, in the appendix, a detailed discussion comparing our
calculation
with that of Takasugi which show explicitly why he obtained the wrong
scaling behaviour.
 
A remark on the calculation of the matrix element is also due.
While we think we have proved the correctness of the analytic
expression for the nuclear operator involved in the $0\nu\beta\beta$
with a composite Majorana neutrino, there still remain some
uncertainty on the actual value of the matrix element ${\cal M}_{FI}$.
As reported in ref.~\cite{hirsch}, ${\cal M}_{F'}$ and 
 ${\cal M}_{GT'}$ are quite sensitive to the value of $m_A$, the cutoff
parameter in the nucleon form factor, and $l_C$, the cutoff parameter 
of the short-range correlations between nucleons.
This is due to the fact that the radial functions appearing in 
${\cal M}_{F'}$ and ${\cal M}_{GT'}$~Eq.(\ref{matrix_list})
are not positive definite 
resulting  in a delicate cancellation within the radial
integrals. This does not happen for  ${\cal M}_{T'}$, ${\cal M}_{GT,N}$
and  ${\cal M}_{FN}$ which are rather stable in the region 
$650\, \hbox{MeV} < m_A < 1.5\, \hbox{GeV} $.
The numerical values used here to get Eq.~(\ref{mfi_num}) 
refer to the values $m_A = 0.85 \, \hbox{GeV}$ and $l_C = 0.7\, \hbox{fm}$.

The sensitivity of some of the nuclear matrix elements 
involved in this calculation together with the fact that 
${\cal M}_{FI}^{(\Delta)}$  is also quite important
 simply indicates that due to the
$\sigma_{\mu\nu}$
coupling appearing in our effective Lagrangian, nuclear 
physics aspects
of the neutrino-less double beta decay 
calculation become more important.
Further investigation is necessary. In particular, 
we believe that the term $I'$ in Eq.~(\ref{q_0integ})
containing $q_0^2$ in the numerator, gives a bigger weight
to the high $q_0$ region and in order to properly account
for the nuclear physics it might be necessary to go beyond
the closure approximation.

Let us now compare our result in Eq.~(\ref{constraint_num}) with those
from high energy experiments. The ZEUS and H1 collaborations (DESY)
have recently published~\cite{zeus,hera} 
results of a direct search of singly produced excited
states in electron-proton collisions
at HERA. They have studied the reaction
$e p \to l^* X $ with the subsequent decay $l^* \to l' V$
where $V= \gamma , Z, W $ (see Fig.~6).
Upper limits for the quantity $\sqrt{|c_{Vl^*e}|^2
+|d_{Vl^*e}|^2}/\Lambda_{\hbox{c}}\times \hbox{Br}^{1/2}(l^* \to l' V)$
are derived\cite{zeus} 
as a function of the excited lepton mass and for the various decay
channels. These experiments were sensitive up to $180$ GeV for $m_{\nu^*}$  and
up to $250$ GeV for $m_{e^*}$ ($m_{q^*}$).
For the purpose of comparing  our  analysis of 
double-beta decay bounds on compositeness with
the high energy bounds, 
we quote here the limit on the $\nu^*$ coupling that the ZEUS
collaboration has 
obtained at the highest accessible mass ($m_{\nu^*} = 180$ GeV):
\begin{equation}
\frac{ \sqrt{|c_{W\nu^*e}|^2
+|d_{W\nu^*e}|^2}}{\Lambda_{\hbox{c}}} \times 
\hbox{Br}^{1/2}(\nu^* \to \nu W)\leq 5 \times 10^{-2} \, \hbox{GeV}^{-1}.
\label{zeus_bound1}
\end{equation}
Let us emphasise that  these limits depend on the branching 
ratios of the decay channel chosen.
For $m_{\nu^*} = 180 $ GeV (the highest accessible mass at the
HERA experiments~\cite{zeus,hera} with $\Lambda_{\hbox{c}} = 1$ TeV,
$Br(\nu^* \to \nu W)=0.61$~\cite{djouadi} and 
$|c_{W\nu^*e}| = |c_{W\nu^*e}|$ one has:
\begin{equation}
|f|< 61. \qquad \qquad \hbox{HERA}
\label{zeus_bound2}
\end{equation}
For the same values of  $m_{\nu^*} = M_N$ and  $\Lambda_{\hbox{c}}$
one obtains from the 
$0\nu\beta\beta$ constraint i.e. Eq.~(\ref{constraint_num}):
\begin{equation}
|f|< 3.84 \qquad \qquad \hbox{\it $0\nu\beta\beta$ }
\end{equation}
We can thus conclude that
the bounds that can be derived from the low-energy
neutrino-less double beta decay are roughly of the same order
of magnitude as those obtained from the direct search of
excited states in high energy experiments.
We also note that, in contrast to  bounds from the direct search
of excited particles, our $0\nu\beta\beta$ constraints on 
$\Lambda_{\hbox{c}} $ and $\vert f  \vert $ 
{\it do not depend } on  any assumptions regarding the 
branching ratios
of the decaying heavy particle. 

Finally, let us conclude by recalling that with respect to the work of 
ref.~\cite{panella,pantrento}, we  have 
improved the calculation of the
phase-space using the exact values of the integrals given in
ref.~\cite{doi}, thus removing the Rosen-Primakoff approximation
for the Fermi functions. This however does not give 
appreciable changes  in the numerical results as shown in Figures
2 and 3. The results appear stable.
                                 
\appendix
\section*{Details of calculation}

Here we show in some detail where the calculation of 
ref.~\cite{takasugi} differs from 
ours and why the author of reference~\cite{takasugi} 
got the wrong scaling behavior for Eq.~(2).
The discrepancy is that the author there used throughout 
an effective four-fermion interaction, i.e. an effective 
Fermi theory. This is clear in view of the absence of W gauge boson 
propagators in his formulae. The use of an effective 
four-fermion interaction of course 
amounts to let $M_W \to \infty$ after having 
introduced the effective Fermi coupling constant $G_F$. Normally
this is  a 
good approximation if there is no other mass scale comparable with or 
larger than $M_W$, as is the case, for instance, in typical 
low energy processes. 
However, in our problem, another mass scale enters the game, namely the 
heavy neutrino mass $M_N$ which we assume to be much greater than $M_W$. It is 
then inconsistent to let $M_W \to \infty$ while keeping $M_N$ finite. 
Nevertheless, this was done in ref.~\cite{takasugi}. 
Since we are to evaluate effects 
due to the heavy mass $M_N$, we must include effects due to 
$M_W$. 
They can not be discarded a priori if we wish to discover the 
correct scaling behavior.

Taksugi's Eq.~(7) in ref.~\cite{takasugi} reads:
\begin{eqnarray}
S_{fi}
&=&i{{G_{eff}^2} \over{(2\pi)^4}} \int dxdy\int dq \, M_N\,
{{\exp{[-iq\cdot (x-y)]}} \over {q^2-M^2_N}}\, 
\langle N_F\mid T(J_\nu^\dagger(x)J_\sigma^\dagger(y)\mid N_I\rangle \,
\times \nonumber \\
&\ & \frac{1}{\sqrt{2}}\, (1-P_{12})\,
t^{\nu\sigma}(E_1,E_2,q^0,\bm{q})\, \exp{\left[i(E_1
x^0+E_2 y^0)\right]}
\end{eqnarray}
\begin{eqnarray}
t^{\nu\sigma}(E_1,E_2,q^0,\bm{q})&=&\bar \psi_S(E_2)
\sigma^{\mu\nu}\sigma^{\rho\sigma}
(\eta_L^{2 }L+\eta_R^{2} R)\psi_S^C(E_1)(q_\mu-E_1  g_{\mu 0})
(q_\rho+E_2 g_{\rho 0})\nonumber \\
G_{eff}&=& \left(\frac{f}{\Lambda_{\hbox{c}}}\right)\, G_F \cos\theta_C 
\qquad \hbox{(homodoublet model)}
\end{eqnarray}
In Eqs.~(A$1$-A$2$) Takasugi is
also neglecting the momenta of the outgoing electrons everywhere except
in the electron wave function (which includes the relativistic Coulomb 
corrections of the nuclear field).
Our Eq.~(8) on the other hand may also be written as:
\begin{eqnarray}
S_{fi}&=&i\,
(G_F\cos\theta_C)^2\,\left(\frac{f}{\Lambda_{\hbox{c}}}\right)^2
\, M_N\,\frac{1}{\sqrt{2}}\, 
(1-P_{12})\,
\int d^4x\,\,d^4y\,\int 
\frac{d^4q}{(2\pi)^4}\,{\exp\left[ -iq \cdot (x-y)\right]
\over {q^2-M^2_N}}\times
\nonumber \\
&& 
M_W^4\, \frac{\langle N_F\mid T(J^\mu_h(x)J^\nu_h(y)\mid N_I\rangle}
{[(q-p_1)^2-M_W^2][(q+p_2)^2-M_W^2]} \,
{\bar \psi}(p_2)\sigma_{\mu\lambda}
\sigma_{\nu\rho}(\eta_L^2 L +\eta_R^2 R)\psi^C(p_1) \times
\nonumber \\
&\ &  \exp\left[ i (p_1\cdot x
+ p_2 \cdot y)\right]\, (q-p_1)^\lambda(q+p_2)^\rho 
\end{eqnarray}
In order 
to compare Eq.~(A$3$) with Eq.~(A$1$) we  neglect, at this stage,
in Eq.~(A$3$), the electron
momenta everywhere but in the wave function (as opposed 
to section III where we did so after having
extracted the four-dimensional delta-function of energy-momentum 
conservation). Hence Eq.~(A$3$) reads :
\begin{eqnarray}
S_{fi}&=&i\, G_{eff}^2 M_N\,
\int d^4x\,\,d^4y\,\int 
\frac{d^4q}{(2\pi)^4}{\exp\left[-iq\cdot (x-y)\right]
\over {q^2-M^2_N+i\epsilon}}\, M_W^4\, 
\frac{\langle N_F\mid T(J^\mu_h(x)J^\nu_h(y)\mid N_I\rangle}
{(q^2-M_W^2+i\epsilon)^2} \times
\nonumber \\
&\ & 
\frac{1}{\sqrt{2}}\, 
(1-P_{12})
t_{\mu\nu}(E_1,E_2,q^0,\bm{q})\, \exp\left[i(E_1
x^0+E_2 y^0) \right]\, .
\end{eqnarray}
Our result  coincides with Eq.~(A1) in the limit $M_W \to \infty$.
We have thus shown that our equations would coincide with those of 
ref.~\cite{takasugi}
if we were to adopt the Fermi's effective theory.

Let us  comment further on the calculation presented in
ref.~\cite{takasugi}.
From Eq.~(A1) Takasugi first performs the $q^0$ integration 
picking up contributions only from the Majorana neutrino pole. 
However, for us, keeping the W propagator is essential.

Let us now continue the calculation from Eq.~(A4), but following the
method employed in ref.~\cite{takasugi}, deviating somewhat  from that
of the present work, but with the exception of postponing the $q^0$
integration. 
Expanding the T-product, inserting a complete set of 
intermediate states between the hadronic current operators,
and integrating out $x_0,y_0$, after some
algebra (exchanging $x \leftrightarrow y $, $q \leftrightarrow
-q$, $\mu \leftrightarrow \nu $ in the second term) 
Eq.~(A$4$) yields :
\begin{eqnarray}
S_{fi}& = &2\pi\, \delta(E_I-E_F-E_1-E_2)\, R_{fi}\nonumber \\
R_{fi} &=& \frac{2G_{eff}^2}{\sqrt{2}} \, M_N\, M_W^4\,\sum_X \,
\int d^3\bm{x}\, d^3\bm{y}\, \frac{d^4q}{(2\pi)^4}
\, \exp{[i\bm{q}\cdot(\bm{x}-\bm{y})]}\,
\langle N_F | J^\mu(\bm{x}) |X \rangle \, 
 \langle X | J^\nu(\bm{y}) |N_I \rangle \, \times\nonumber\\
&\ & (1-P_{12})\,\frac{
t_{\mu\nu}(q;E_1,E_2)}{\left(q_0+\Delta_X^{(1)}-i\epsilon\right)
(q^2-M_N^2+i\epsilon)(q^2-M_W^2+i\epsilon)^2}
\end{eqnarray} 
where $\Delta_X^{(1)} = E_X -E_F-E_1 = E_X - 1/2(E_I+E_F)-1/2(E_1-E_2)
\approx E_X -1/2(M_I+M_F)$.
As in section III of this work, $E_X$ is replaced by an average
excitation energy $\langle E_X \rangle $ (closure approximation):
\begin{equation}
\Delta_X^{(1,2)} \to \Delta = \langle E_X \rangle - 
\frac{1}{2}(M_I+M_F) \approx 10 \, \hbox{Mev.}
\end{equation}
Neglecting the electronic energies also in the numerator 
(in the tensor $t_{\mu\nu}$) and using the spinor identity 
 given in Eq.~(\ref{spin_iden}) we obtain :
\begin{eqnarray}
R_{fi} &=& \frac{4G_{eff}^2}{\sqrt{2}} \, M_N\, M_W^4\,
\int d^3\bm{x}\, d^3\bm{y}\, \frac{d^4q}{(2\pi)^4}
\, \exp{[i\bm{q}\cdot(\bm{x}-\bm{y})]}\,\langle N_F | J^\mu(\bm{x}) 
J^\nu(\bm{y}) |N_I \rangle\times\nonumber\\
&\ &  \frac{\eta_{\mu\nu}q^2 - 
q_\mu q_\nu}{(q_0+\Delta-i\epsilon)
(q^2-M_N^2+i\epsilon)(q^2-M_W^2+i\epsilon)^2}\times \bar{\psi}(2)
(\eta_L^2L +\eta_R^2R)\psi^C(1)
\end{eqnarray}
Now we make use of the non-relativistic impulse approximations for 
the hadronic charged current, just as in section III (inclusion
of nuclear form factor etc.) and neglect
terms in $q_0q_i$ since we only want to consider $0^+ \to 0^+ $
transitions.
We find:
\begin{eqnarray}
R_{fi}(0^+ \to 0^+) &=& \frac{4G_{eff}^2}{\sqrt{2}} \, \bar{\psi}(2)
(\eta_L^2L +\eta_R^2R)\psi^C(1)\, M_N\, M_W^4\,
\int \, \frac{d^3\bm{q}}{(2\pi)^3} \, f_A^2(\bm{q}^2)\,\times\nonumber\\
&\, & \sum_{k,l}\, 
\langle N_F | \exp{(i\bm{q}\cdot{\bm{r}}_{kl})}
\tau_+^{(k)}\,\tau_+^{(l)}\biggl\{
- g_V^2\,\bm{q}^2 \, {\cal I}(\bm{q}^2)\, 
\nonumber \\
&\, &
+ g_A^2\left[ -\eta^{ij}\bm{q}^2\, -\,q^iq^j
\right]\, {\cal I}(\bm{q}^2)\, \sigma^{(k)}_i\,  \sigma^{(l)}_j
- g_A^2
\sigma^{(k)}\cdot\sigma^{(l)}
\, {\cal I'}(\bm{q}^2)\, \biggr\} |N_I \rangle
\end{eqnarray}
where we have defined:
\begin{eqnarray}
{\cal I}(\bm{q}^2) & = &\,\frac{\partial}{\partial \omega_W^2} 
\, {\cal J}(\bm{q}^2) \cr
{\cal J}(\bm{q}^2) & = &  \int \frac{dq_0}{2\pi}\, 
\biggl[\frac{1}{q_0-\Delta +i\epsilon} 
\biggr]\, \frac{1}
{(q_0^2-\omega_N^2 +i\epsilon)
\,(q_0^2-\omega_W^2 +i\epsilon)} \cr
{\cal I'}(\bm{q}^2) & = &\,
\frac{\partial}{\partial \omega_W^2} \, J'(\bm{q}^2) \cr
{\cal J'}(\bm{q}^2) & = &  \int \frac{dq_0}{2\pi} \,
\biggl[ \frac{1}{q_0-\Delta +i\epsilon}
\bigg]\, \frac{q_0^2}
{(q_0^2-\omega_N^2 +i\epsilon)
\,(q_0^2-\omega_W^2 +i\epsilon)}
\label{q_0integ_new}
\end{eqnarray}
with $\omega_N$ and $\omega_W$ are the same as in section III.

Eq.~(A8) is found to be equivalent to Eq.~(\ref{tfi_last}) because of the 
identity:
\begin{equation}
\int \frac{dq_0}{2\pi}\left[ \frac{1}{q_0+\Delta-i\epsilon}\right] f(q_0^2)=
-\frac{1}{2}
\int \frac{dq_0}{2\pi i} \left[
\frac{2 i \Delta}{q_0^2-\Delta^2+i\epsilon}\right] f(q_0^2)
\end{equation}
which 
implies that the integrals ${\cal I}, {\cal I'}$ are proportional to
$ I ,I' $ defined in Eq.~(\ref{q_0integ}), section IV, namely :
${\cal I}  =  ({i}/{2}) \,I \, ;\, {\cal I'}  =   ({i}/{2}) \, I'$.
The fact Eq.~(A8) coincides with Eq.~(\ref{tfi_last}) concludes 
our proof that including the W boson propagators in the
calculation of ref.~\cite{takasugi} gives the correct scaling
behaviour as found in the present work 
(and also found in ref.~\cite{panella}).


%
%

%
%
\begin{table}
\caption{Nuclear structure matrix elements for $^{76}$Ge
as reported by the authors of ref.~\protect\cite{hirsch}.}
\label{}
\begin{tabular}{ccccc}
${\cal M}_{GT,N}$ & ${\cal M}_{FN}$ &${\cal M}_{GT'}$ & ${\cal M}_{F'}$&
${\cal M}_{T'}$\cr
1.13$\times 10^{-1}$ &$-$4.07$\times 10^{-2}$ &$-$7.70$\times 10^{-3}$ &
3.06$\times 10^{-3}$ &$-$3.09$\times 10^{-3}$\cr
\end{tabular}
\end{table}

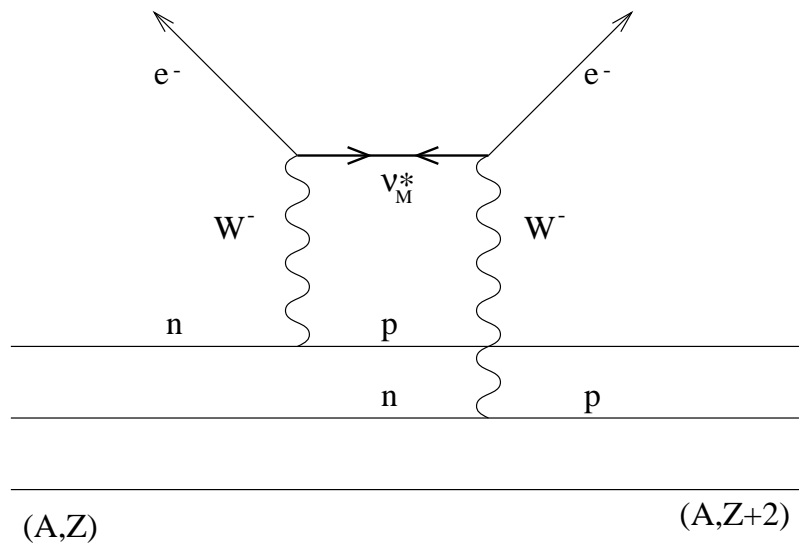
\begin{figure}[htbp]
  \begin{center}
    \leavevmode
    \caption{Neutrinoless double beta decay ($\Delta L = +2$ process)
             mediated by a  heavy composite Majorana neutrino.}
    \label{fig:1}
  \end{center}
\end{figure}

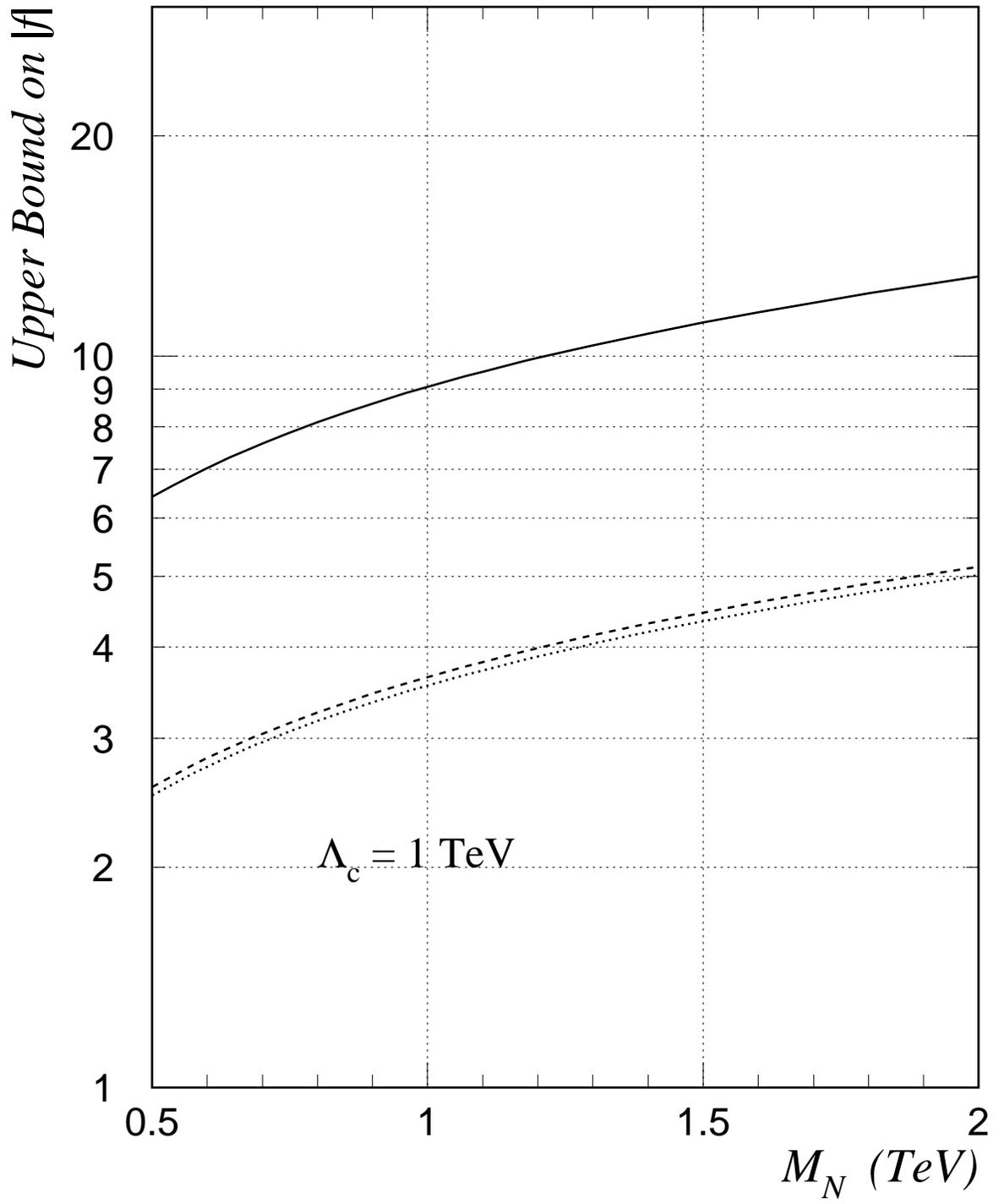
\begin{figure}[htbp]
  \begin{center}
    \leavevmode
    \caption{New bounds on the parameter $|f|$ (for
             $\Lambda_{\protect\hbox{c}} = 1\protect\hbox{TeV}$) from 
             $0\nu\beta\beta $
             (solid line) compared with the estimate, based on a
             upper bound of the nuclear matrix element 
             ($|{\cal M}_{FI} | < 0.34 $), given in
             ref.~\protect\cite{panella,pantrento} (dashed line).
             The dotted line is the calculation of
             ref.~\protect\cite{panella,pantrento} augmented by the
             exact phase space calculation (no Rosen-Primakoff
             approximation for the Fermi functions).}
    \label{fig:2}
  \end{center}
\end{figure}

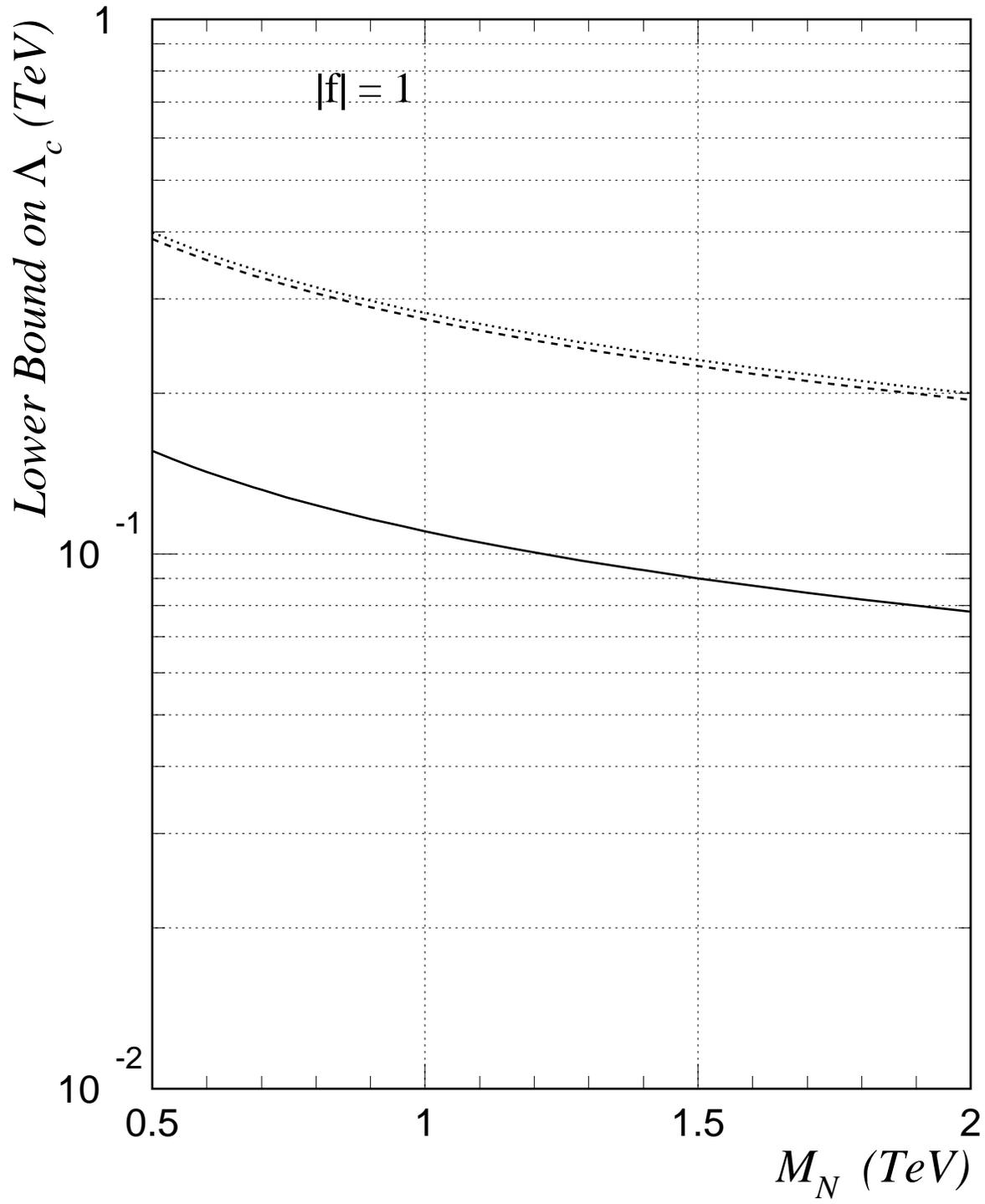
\begin{figure}[htbp]
  \begin{center}
    \leavevmode
    \caption{New bounds on the parameter $\Lambda_{\protect\hbox{c}}$ 
             (for $|f|$=1)
             from $0\nu\beta\beta $ (solid line) compared with the 
             estimate given in
             ref.~\protect\cite{panella,pantrento} (dashed line).
             The dotted line is the calculation of
             ref.~\protect\cite{panella,pantrento} augmented by the
             exact phase space calculation (no Rosen-Primakoff
             approximation for the Fermi functions).}
    \label{fig:3}
  \end{center}
\end{figure}


\begin{references}
\bibitem{trento}
                Proceedings of the International Workshop
                ``{\it Double Beta Decay and Related Topics}''
                held at the European Centre for Theoretical Studies
                (ECT$^*$), Trento, Italy, April 24-May 5, 1995. Ed. 
                H.~V.~Klapdor-Kleingrothaus and 
                S.~Stoica, World Scientific, Singapore, 1996.

\bibitem{mohapatra} 
                For a review of the possible mechanisms of 
                $0\nu\beta\beta$ see R.~N.~Mohapatra in the 
                proceedings of the Trento workshop, c.f.~\cite{trento}.

\bibitem{hirsch1}
                M.~Hirsch, H.~V.~Klapdor-Kleingrothaus and S.~G.~Kovalenko,
                Phys. Lett. B {\bf 352} (1995), Phys. Rev. Lett. {\bf 75} 
                (1995), 17;
                Phys. Rev. D {\bf 53} (1996) 1329.

\bibitem{hirsch2}
                M.~Hirsch, H.~V.~Klapdor-Kleingrothaus and O.~Panella,
                Phys. Lett. B {\bf 374}, 7-12, 1996; 
                e-Print archive: hep-ph 9602306. 

\bibitem{hirsch3}
                M.~Hirsch, H.~V.~Klapdor-Kleingrothaus and
                S.~G.~Kovalenko, Phys. Rev. D {\bf 54}, 4207-4210, 1996,
                e-Print archive: hep-ph/9603213; Phys. Lett. B {\bf
                378}, 17-22, 1996, e-Print archive: hep-ph/9602305.
 

\bibitem{burgess}
                C.~P.~Burgess and J.~M.~Cline, 
                Phys. Rev. D {\bf 49}, 5925-5944, 1994;
                M.~Hirsch, H.~V.~Klapdor-Kleingrothaus and
                S.~G.~Kovalenko, Phys. Lett. B {\bf 372}, 8-14, 1996;
                e-Print archive: hep-ph/9511227.

\bibitem{panella}
                O.~Panella and Y.~N.~Srivastava, 
                Phys. Rev. D {\bf 52} 5308-5313, 1995.
\bibitem{HM}
                B.~Maier et al. in the proceedings of the 
                Trento workshop~\cite{trento}.

\bibitem{takasugi}
                E.~Takasugi, Prog. Theor. Phys. {\bf 94}, 1097-1104, 1995, 
                e-Print archive: hep-ph/9506379; 
                Also published in ref.~\cite{trento}

\bibitem{CDF}   F.~Abe et al. (CDF Coll.), 
                Phys. Rev. Lett. {\bf 77}, 438-443, 1996,
                FERMILAB-PUB-96/020-E, e-Print archive: hep-ex 9601008.

\bibitem{preons}
       H.Terazawa, Y.~Chikashige and K.~Akama, Phys Rev D {\bf 15}
       480 (1977);
       H.~Harari, Phys. Lett. B {\bf 86}, 83 (1979);
       H.~Fritzsch, G.~Mandelbaum, Phys. Lett. B {\bf 102}, 
       319 (1981); O.~Greenberg and J.~Schuler, {\it ibid} {\bf 99}
       339 (1981); R.~Barbieri, R.~N.~Mohapatra, A.~Masiero, {\it ibid}
       {105} 369 (1981).\\ 
       For further references see for example:
       H.~Harari, Phys. Rept. {\bf 104}, 159 (1984);
       I.~A.~D'Souza and C.~S.~Kalman, {\it Preons, Models
       of Leptons, Quarks and Gauge Bosons as Composite Objects},
       World Scientific Publishing Co. Singapore, 1992.    

\bibitem{pantrento}
                O.~Panella in the proceedings of the Trento workshop
                c.f.~\cite{trento}.




\bibitem{pdg}
                R.~M.~Barnett {\it et al.}
                (Particle Data Group), {\it Review of
                Particle Properties};
                Phys. Rev. D {\bf 54} 1, (1996).
\bibitem{cab}
                N.~Cabibbo, L.~Maiani and Y.~Srivastava,
                Phys. Lett. B {\bf 139} 459 (1984).


\bibitem{ruj}   
                A.~De~Rujula, L.~Maiani and R.~Petronzio,
                Phys. Lett. B {\bf 140}, 253 (1984).

\bibitem{baur}
                U.~Baur, I.~Hinchliffe and D. Zeppenfeld,
                Int. J. Mod. Phys. A {\bf 2} 1285, (1987).
       
\bibitem{pan}
                G.~Pancheri and Y.N.~Srivastava,
                Phy. Lett. B {\bf 146}, 87 (1984).


\bibitem{lipkin}
                H.~J.~Lipkin, A.~de~Shalit and I.~Talmi,
                Nuovo Cimento, {\bf 2}, 773, 1955.\\
                See also:
                P.~Ring and P.~Schuck 
                ``{\it The Nuclear Many body Problem}''
                chap. XI pag. 451-456.
                Springer \& Verlag, New York, 1980.  

\bibitem{deshalit}
                See for example: A.~de Shalit and 
                H.~Feshbach, {\it Theoretical Nuclear Physics} Vol. I, 
                chap. IX, pag. 840-861, and 881-885, 
                John Wiley \& and Sons, New York, 1974.

\bibitem{hirsch}
                M. Hirsch, H.V. Klapdor-Kleingrothaus
                and  S.G. Kovalenko, Phys. Rev. D {\bf 53}, 1329-1348,
                1996, e-Print archive: hep-ph/9502385.

\bibitem{doi}   
                M.~Doi, T.~Kotani and E.~Takasugi,
                Prog. Theor. Phys. Supplement {\bf 83}, 1 (1985).
\bibitem{HMnew} 
 	M. G\"unther et al. (Heidelberg-Moscow Collaboration),
	Phys. Rev. D {\bf 55}, 54 (1997).
\bibitem{zeus}
       M.~Derrik et al. (ZEUS Coll.) {\it Z. Phys.} C 
       {\bf 65}, 627 (1995).  
\bibitem{hera}   
       F.~Raupach, (H1 Collab.) in the
       {\it Proceedings of the International Europhysics 
       Conference on High Energy Physics}, Marseilles,
       France 22-28 July 1993. Editors: J.~Carr and M.~Perrottet.
       Editions Fronti\`ers, Gif-Sur-Yvette, France, 1994.
\bibitem{djouadi}
        A.~Djouadi, J.~Ng and T.~G.~Rizzo, SLAC preprint
        SLAC-PUB-95-6772, electronic archive hep-ph/9504209;
        To appear as a chapter in: ``{\it Electroweak
	Symmetry Breaking and Beyond the Standard Model} '', 
	edited by T. Barklow, S.Dawson, H.E. Haber and 
	S. Siegrist, World Scientific.
\end{references}
\end{document}